\documentclass[conference]{IEEEtran}
\IEEEoverridecommandlockouts
\usepackage[T1]{fontenc}
\usepackage[utf8]{inputenc}
\usepackage{microtype}
\usepackage{graphicx}
\usepackage{booktabs}
\usepackage{multirow}
\usepackage{amsmath}
\usepackage{xspace}
\usepackage{xcolor}
\usepackage{url}
\usepackage[hidelinks]{hyperref}
\usepackage{listings}
\usepackage{comment}
\usepackage{balance}  % balance last-page columns

\newcommand{\omrag}{\textsc{OM-RAG}\xspace}
\newcommand{\czero}{\texttt{C-zero}\xspace}
\newcommand{\cchunk}{\texttt{C-chunk}\xspace}
\newcommand{\cdockg}{\texttt{C-doc-kg}\xspace}

\newcommand{\etal}{\textit{et al.}\xspace}
\definecolor{lightgreen}{RGB}{220,245,220}

\title{Leveraging Resolved Incident History for LLM-Assisted Software Bug Diagnosis}

\author{Boyuan Guan$^{1}$, Hailu Xu$^{2}$, and Jamie Rogers$^{3}$% <-this % stops a space
\thanks{$^{1}$Boyuan Guan is with the Geographic Information System (GIS) Center,
        Florida International University, FL 33199, USA.
        {Email: \tt\small bguan@fiu.edu}}%
\thanks{$^{2}$Hailu Xu is with the Department of Computer Engineering and Computer Science, California State University,
        Long Beach, CA 90840, USA.
        {Email: \tt\small hailu.xu@csulb.edu}}%
\thanks{$^{3}$Jamie Rogers is with the FIU Libraries, Florida International University,
        FL 33199, USA.
        {Email: \tt\small rogersj@fiu.edu}}%
}

\begin{document}

\maketitle
\thispagestyle{empty}
\pagestyle{empty}
% ─────────────────────────────────────────────────────────────────────────────
\begin{abstract}
Effective software bug diagnosis requires two ingredients: the right knowledge source (operational failure history, not just system documentation) and the right retrieval structure (structured records, not unstructured chunks). Current retrieval-augmented generation (RAG) approaches fall short on one or both dimensions. We propose Operational Memory RAG (OM-RAG), which indexes resolved issues as structured symptom--root-cause--resolution triples and retrieves the most similar historical precedent via single-hop embedding. OM-RAG powers a purpose-built large language model (LLM) administrator that has operated a production Dataverse instance for over six months, operationalizing the Knowledge component of the Monitor--Analyze--Plan--Execute--Knowledge (MAPE-K) feedback loop with the system's own resolved failure history. In a controlled four-configuration ablation, LLM-based judging covers all 1,172 issues and manual verification a 123-issue subset. OM-RAG achieves Diagnosis Accuracy of 0.931 and Fix Correctness of 0.809, outperforming chunk-based retrieval ($+$186\%) and a documentation-anchored concept graph ($+$77\%). Model-free retrieval metrics (mean similarity 0.880) independently corroborate the ranking.

\end{abstract}


% ─────────────────────────────────────────────────────────────────────────────
\section{Introduction}
\label{sec:intro}
Modern RAG systems for software operations fail not because they retrieve irrelevant information, but because they retrieve the \textit{wrong type} of knowledge~\cite{barnett2024seven}. An LLM agent with access to complete system documentation possesses rich semantic knowledge of how a system is \textit{designed} to work, yet it often fails to diagnose real production failures. The missing ingredient is not more documentation. It is \textit{episodic operational memory} \cite{tulving1972episodic}, the structured recall of how similar failures have previously manifested and been resolved in a specific system.

This gap mirrors a familiar phenomenon in human expertise. A medical student may memorize every physiology textbook yet still underperform an experienced clinician \cite{schmidt1990encapsulation}. The difference is accumulated case experience: recalling how similar symptoms have resolved in practice.

Consider a standard RAG agent diagnosing a bug in Dataverse, a university research data repository, where DOI minting silently fails after a configuration change. The agent retrieves documentation about DOI integration, EZID authentication, and dataset publication. The retrieved content is accurate and grounded, but it does not explain why a specific configuration in \texttt{EZIDServiceBean} disables minting. The agent therefore produces a plausible diagnosis that is factually wrong.

We term this the \textit{\textbf{retrieved but wrong}} failure mode. It arises from two mismatches. First, a \textit{source mismatch}: the system retrieves documentation describing intended behavior rather than historical failure patterns. Second, a \textit{structure mismatch}: even when retrieving from resolved issues, standard text chunking fragments the causal chain across chunks~\cite{barnett2024seven}. The LLM may retrieve symptoms without causes, or causes without resolutions. Effective diagnosis therefore requires both operational knowledge and a structure that preserves the symptom--root cause--resolution relationship.

The MAPE-K feedback loop~\cite{kephart2003vision} (Monitor, Analyze, Plan, Execute, Knowledge) is the standard architectural pattern for self-adaptive systems. The Knowledge (K) component serves as the shared context for all loop functions. Recent surveys~\cite{li2025genaisas} identify a key open challenge: current MAPE-K implementations populate K with static theoretical models rather than dynamic operational records accumulated during system execution.

OM-RAG directly addresses this gap as the diagnostic knowledge component of \textbf{DIVA}\footnote{\url{https://dataverse.fiu.edu/ai/}} (Dataverse Intelligent Virtual Assistant), an LLM-based AI administrator purpose-built by the authors for the FIU Dataverse production instance. Since October 2025, DIVA has managed the instance through real operational incidents (i.e., Solr index corruption, file-upload outages, multi-component cascading faults) accumulating exactly the episodic operational memory that OM-RAG formalizes and retrieves.

This paper makes the following contributions:

\begin{enumerate}
    \item \textbf{Two-dimensional knowledge analysis}: We show that diagnostic performance depends on both the knowledge source (theoretical vs.\ operational) and retrieval structure (unstructured vs.\ structured), and that combining operational knowledge with structured retrieval yields the best results.
    
    \item \textbf{OM-RAG architecture}: We present OM-RAG, which retrieves structured operational cases rather than documentation fragments, preserving the causal relationships required for diagnosis.
    
    \item \textbf{Public benchmark}: We construct a benchmark of 1,172 resolved issues from the \texttt{IQSS/dataverse} repository with verified ground truth, providing an open testbed for operational-memory-based diagnosis.
    
    \item \textbf{Production validation}: OM-RAG powers DIVA, an LLM agent that has administered the FIU Dataverse production instance for over six months, demonstrating the practicality of operational memory in real-world diagnosis.
\end{enumerate}

Section~\ref{sec:setup} reports the full ablation: OM-RAG achieves Diagnosis Accuracy of 0.931, a $+186\%$ improvement over chunk-based retrieval and $+77\%$ over a documentation-anchored knowledge graph, independently corroborated by model-free retrieval metrics.

% ─────────────────────────────────────────────────────────────────────────────
\section{Background and Problem Framing}
\label{sec:background}
\subsection{The Diagnosis Gap in RAG Systems}

Existing approaches implicitly assume that improving retrieval quality---through better embeddings or larger corpora---will improve diagnostic performance. However, diagnosis is not purely a semantic matching task: it requires identifying the causal pattern underlying a failure, which depends on (1) retrieving from operational failure history rather than design-time specifications, and (2) preserving the causal mapping between symptoms, root causes, and resolutions.

\subsection{Two Dimensions of Knowledge for Diagnosis}

We characterize diagnostic knowledge along two dimensions: \emph{source} and \emph{structure}.

The source dimension distinguishes theoretical knowledge (e.g., documentation and specifications) from operational knowledge (e.g., resolved incidents and failure history). The former describes intended behavior, whereas the latter captures how systems actually fail. Diagnosis depends primarily on operational knowledge.

The structure dimension distinguishes unstructured text from structured records. Unstructured chunking fragments the mapping between symptoms, root causes, and resolutions, forcing the model to reconstruct these relationships during generation. Structured representations preserve this causal mapping explicitly and are therefore more suitable for diagnosis.

\subsection{Operational Memory as the Missing Layer}

We argue that the missing component for LLM-based diagnosis is \emph{operational memory}: a structured repository of how a system has failed and been repaired.

This idea is supported by three perspectives. Cognitive science distinguishes semantic memory from episodic memory, where expert diagnosis relies heavily on recalling past experiences~\cite{tulving1972episodic}. Case-Based Reasoning (CBR) similarly solves new problems by retrieving structured prior cases~\cite{aamodt1994case}. In autonomic systems, the Knowledge ($K$) in MAPE-K is often static, while diagnosis requires knowledge accumulated from actual failures.

These perspectives point to a common principle: diagnosis should retrieve structured operational cases rather than isolated text fragments. OM-RAG therefore introduces not a new retrieval algorithm, but a new retrieval target---operational memory encoded as structured diagnostic cases.

% ─────────────────────────────────────────────────────────────────────────────
\section{The OM-RAG Architecture}
\label{sec:architecture}

\begin{figure*}[t]
  \centering
  \includegraphics[width=0.7\linewidth]{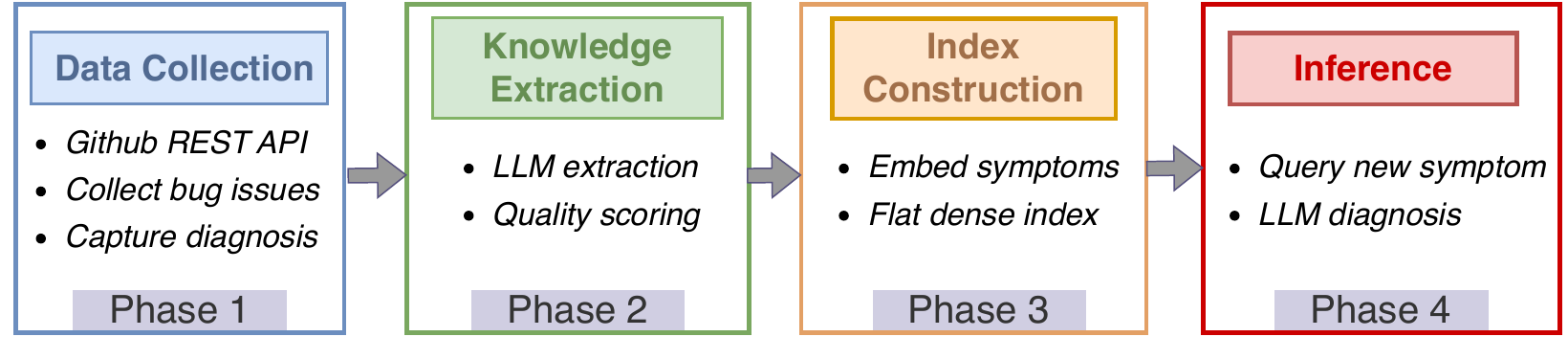}
  \caption{The four-phase \omrag pipeline. Phase~1 collects
  GitHub issues with comment threads. Phase~2 extracts structured triples via LLM and
  quality-filters to records. Phase~3 embeds symptoms into a flat
  dense index. Phase~4 retrieves the top-$k$ most similar triples for a new
  symptom and prompts the LLM to generate a grounded diagnosis.}
  \label{fig:pipeline}
  \end{figure*}
  
  The \omrag system implements a four-phase pipeline that transforms historical bug resolution data into actionable diagnostic knowledge. As illustrated in Figure~\ref{fig:pipeline}, these phases are: (1) Data Collection, which harvests closed bug reports and their resolution threads from the system's issue tracker; (2) Knowledge Extraction, which uses LLMs to distill each incident into structured symptom-root cause-resolution triples with quality scoring; (3) Index Construction, which builds a flat, embedding-based retrieval index optimized for single-hop diagnostic lookup; and (4) Inference, which retrieves historically similar incidents and prompts the LLM to diagnose new failures by reasoning from these precedents. This architecture operationalizes the knowledge component
  $K$ in the MAPE-K loop by grounding the Analyze phase in the system's own resolved failure history rather than design-time specifications alone. The resulting episodic operational memory constitutes one track within a broader multi-track knowledge architecture for governed agentic AI~\cite{guan2026dualhelix,guan2026envismart}, providing the historical grounding necessary to stabilize LLM-driven system administration.

\begin{comment}
  \begin{figure}[t]
  \centering
  \includegraphics[width=0.65\linewidth]{figs_SMC/OM-RAG.pdf}
  \caption{OM-RAG as the episodic Knowledge component in the MAPE-K feedback loop.
  Solid arrows trace the MAPE-K control cycle; dashed arrows show knowledge retrieval
  from the operational memory KB during Analyze and Plan; the dotted arrow represents
  the Retain step that feeds newly resolved incidents back into $K$.}
  \label{fig:mapek}
  \end{figure}
\end{comment}
  
  \subsection{Phase 1: Data Collection}
  
  We use the GitHub REST API to collect all closed issues labeled ``Type: Bug'' from the \texttt{IQSS/dataverse} repository, yielding 2,283 bug reports. Each issue is enriched with its complete comment thread (up to 100 comments per issue) to capture the diagnostic discourse between reporters and developers. After filtering for issues with at least two substantive comments, 1,642 enriched issues remain.
  
  \subsection{Phase 2: Knowledge Extraction}
  
  For each enriched issue, we prompt an LLM (\texttt{gpt-4o-mini}) to extract a structured triple and assign a quality score:
  
  \begin{itemize}
      \item \textbf{symptom}: The observable manifestation as described by the reporter
      \item \textbf{root\_cause}: The underlying cause as identified in the resolution discussion
      \item \textbf{resolution}: The fix applied or the configuration change that resolved it
      \item \textbf{quality\_score} $\in [1, 5]$: How clearly the issue defines all three fields
  \end{itemize}
  
  Issues with \texttt{quality\_score} $< 3$ are excluded. Of the 1,642 enriched issues, 1,641 are successfully extracted; after quality filtering, 1,172 issues with complete, unambiguous ground truth form the evaluation benchmark. The full KB contains 1,457 structured triples from $\sim$1,396 unique issues (some issues produce multiple root-cause patterns).
  
  \subsection{Phase 3: Index Construction}
  
  The operational memory index is constructed by concatenating the \textit{symptom} and \textit{root\_cause} fields of each extracted triple into a single symptom-RC string, then embedding each string using OpenAI \texttt{text-embedding-3-small} (1536 dimensions). The embeddings are stored in a SQLite database alongside the full triple and issue metadata. At inference, the top-$k$ ($k=3$) most similar records are retrieved via cosine similarity.
  
   The index deliberately avoids cross-incident graph structure. The retrieval sub-problem in bug diagnosis is structurally single-hop: the needed knowledge is concentrated in the single best-matching historical incident record, and no cross-document synthesis is required. Adding a graph layer introduces path-traversal indirection (symptom $\to$ concept $\to$ RC) that adds noise without improving precision on single-hop lookups. Section~\ref{sec:setup} empirically confirms this: adding a concept-layer graph reduces Diagnosis Accuracy by 43\% relative to \omrag.

  \subsection{Phase 4: Inference}
  At inference time, the \omrag agent follows this procedure:
  
  \begin{enumerate}
  \item Embed the new issue's title + body using the same encoder.
  \item Retrieve the top-3 most similar symptom-RC records from the index.
  \item Build a prompt containing: (a) the retrieved records as ``historical precedents'',
    (b) the current issue description, and (c) a structured diagnosis request.
  \item Call the LLM (e.g., \texttt{gpt-4o-mini}, \texttt{gpt-5.2}) to
    generate a structured diagnosis: root cause identification, fix recommendation,
    and confidence level.
  \end{enumerate}
  
  The prompt instructs the LLM to explicitly reason from the retrieved precedents
  (``the most similar past issue had this root cause; does this pattern apply here?'')
  rather than from generic system knowledge.

% ─────────────────────────────────────────────────────────────────────────────
\section{Evaluation Results}
\label{sec:setup}
\subsection{Testbed and Deployment}

Florida International University's research data repository is built on the open-source Dataverse platform, developed and supported by Harvard University's Institute for Quantitative Social Science (IQSS). As an active deployment of a complex, multi-component system (Java/Payara application server, PostgreSQL, Apache Solr, Shibboleth SSO), Dataverse has an operational history spanning 10+ years and 15,000+ GitHub issues.
This system is not merely a benchmark target: it is the production environment in which DIVA operates. The \omrag knowledge base is built from the same issue tracker DIVA consults during live operations; each resolved incident is ingested back into the index, making the benchmark a direct reflection of real operational history.

\subsection{Benchmark Construction}

The benchmark applies the Phase~1--2 pipeline of Section~\ref{sec:architecture}
(2,283 bug-labeled issues $\rightarrow$ 1,642 comment-enriched $\rightarrow$ 1,641 extracted
$\rightarrow$ 1,172 retained at \texttt{quality\_score} $\geq 3$). The same LLM extractor
additionally categorizes each issue into 13 primary domain buckets (\textit{api},
\textit{authentication}, \textit{database}, \textit{deployment}, \textit{doi-pid},
\textit{email-notification}, \textit{file-upload}, \textit{harvesting}, \textit{metadata},
\textit{permissions}, \textit{publishing}, \textit{search-solr}, \textit{ui-rendering}),
with a small residual of 13 issues receiving unrecognized labels (\textit{other},
\textit{testing}, \textit{dependencies}).

The resulting 1,172 issues form the evaluation benchmark. The operational memory KB is built from all 1,457 extracted triples. At inference time, each benchmark issue's own extracted record is excluded from the index to prevent data leakage.
Ground truth triples are derived from developer-authored resolution discussions (commit references, maintainer confirmations, and accepted fix descriptions) in the GitHub issue threads; they represent actual human-verified resolutions, not synthesized content.

\subsection{System Configurations}

\begin{table}[t]
\centering
\caption{Evaluated System Configurations}
\label{tab:configs}
%\footnotesize
\small
\begin{tabular}{@{}llll@{}}
\toprule
\textbf{Config} & \textbf{Source} & \textbf{Representation} & \textbf{Retrieval} \\
\midrule
\czero & None & --- & None \\
\cchunk & Issues & 1,147 chunks & FAISS \\
\cdockg & Doc-anchored & Graph & Query$\rightarrow$RC \\
\omrag & Issues & 1,457 triples & Embed top-k \\
\bottomrule
\end{tabular}
\end{table}

We evaluate four system configurations (see Table~\ref{tab:configs}).
\cchunk uses unstructured chunks from resolved issues; \omrag uses structured triples
from the same issues---isolating \emph{representation}.
\cdockg derives its concept taxonomy from official Dataverse documentation, with
resolved issues mapped to those documentation concepts; comparing \cdockg vs.\ \omrag
primarily targets \emph{knowledge source} (documentation-anchored vs.\ incident-indexed).
All four configurations use the same inference LLM (\texttt{gpt-5.2}) to ensure
that observed differences are attributable solely to knowledge base and retrieval
architecture.\footnote{Results are specific to \texttt{gpt-5.2} and
\texttt{claude-sonnet-4-6}; performance may vary with different model versions.}

\begin{table}[t]
\centering
\caption{Knowledge Type Classification of Evaluated Configurations}
\label{tab:knowledge-types}
\small
\resizebox{\columnwidth}{!}{
\begin{tabular}{@{}lllll@{}}
\toprule
 & \textbf{\czero} & \textbf{\cchunk} & \textbf{\cdockg} & \textbf{\omrag} \\
\midrule
\textbf{KB Source} & Parametric & Issues & Doc-anchored & Issues \\
\textbf{Type} & --- & Operational & Theoretical & Operational \\
\textbf{Memory} & Semantic & Episodic & Semantic & Episodic \\
\textbf{Structure} & None & Chunks & Graph & Triples \\
\bottomrule
\end{tabular}
}
\end{table}

Table~\ref{tab:knowledge-types} maps these dimensions to the four system configurations we evaluate.
These comparisons probe each dimension: \cchunk vs.\ \omrag cleanly isolates structure (both use operational issues); \cdockg vs.\ \omrag primarily targets knowledge source, though \cdockg's graph traversal also differs from \omrag's flat retrieval---an inherent property of documentation-anchored architectures, where the graph \emph{is} the mechanism through which documentation knowledge is organized.

\begin{table*}[t]
\centering
\caption{Decomposed evaluation scores ($n{=}1{,}172$, full benchmark).
DA=Diagnosis Accuracy, FC=Fix Correctness, G=Groundedness, S=Specificity.
Mean $\pm$ half-width of 95\% bootstrap CI; scale 0/0.5/1.0 per dimension, max total = 4.0.}
\label{tab:main-results}
\small
\setlength{\tabcolsep}{4pt}
\renewcommand{\arraystretch}{1.2}
\begin{tabular}{@{}lccccc@{}}
\toprule
\textbf{Config} & \textbf{DA} & \textbf{FC} & \textbf{G} & \textbf{S} & \textbf{Total} \\
\midrule
\czero  & $0.238{\pm}.017$ & $0.184{\pm}.015$ & $0.429{\pm}.011$ & $0.453{\pm}.009$ & $1.303{\pm}.038$ \\
\cchunk & $0.325{\pm}.022$ & $0.187{\pm}.016$ & $0.865{\pm}.013$ & $0.365{\pm}.013$ & $1.742{\pm}.048$ \\
\cdockg & $0.526{\pm}.024$ & $0.373{\pm}.018$ & $0.684{\pm}.014$ & $0.501{\pm}.005$ & $2.084{\pm}.051$ \\
\omrag  & \colorbox{lightgreen}{$\mathbf{0.931}{\pm}.012$} & \colorbox{lightgreen}{$\mathbf{0.809}{\pm}.015$} & \colorbox{lightgreen}{$\mathbf{0.965}{\pm}.007$} & \colorbox{lightgreen}{$\mathbf{0.562}{\pm}.010$} & \colorbox{lightgreen}{$\mathbf{3.266}{\pm}.033$} \\
\bottomrule
\end{tabular}
%\vspace{-0.2in}
\end{table*}

\subsection{Evaluation Metrics}

\textbf{Decomposed 4-dimension rubric.} Following RAGAS~\cite{es2023ragas}'s principle of decomposed independent scoring, we adapt the framework for the bug diagnosis task.
RAGAS defines Faithfulness (are claims grounded in retrieved context?) and Answer Relevance (does the answer address the question?).
We retain Faithfulness as Groundedness (G), replace Answer Relevance with two task-specific correctness checks against verified ground truth -- Diagnosis Accuracy (DA) and Fix Correctness (FC), and add Specificity (S) to capture diagnostic precision not addressed by RAGAS.
RAGAS's retrieval-side metrics (Context Precision/Recall) are replaced by our model-free \texttt{flat\_rc\_score} and \texttt{gt\_keyword\_hit\_rate} (defined below).
Each dimension is scored $\{0, 0.5, 1.0\}$:

\begin{itemize}
    \item \textbf{DA} (Diagnosis Accuracy): Does the response correctly identify the root cause stated in the ground truth?
    \item \textbf{FC} (Fix Correctness): Do the proposed resolution steps substantially overlap with the ground truth fix?
    \item \textbf{G} (Groundedness): Are the key claims traceable to the provided context rather than hallucinated?
    \item \textbf{S} (Specificity): Does the response name concrete artifacts (file names, class names, config keys, commands)?
\end{itemize}

Total score = DA + FC + G + S $\in [0, 4.0]$. Scoring is performed by \texttt{claude-sonnet-4-6} as the primary LLM judge, treated as an approximate oracle whose rankings are corroborated by model-free metrics (Section~\ref{subsec:scale}).

\textbf{Model-free metrics.} Two corroborating metrics require no LLM judge:
\texttt{flat\_rc\_score}: cosine similarity between the query embedding and the retrieved RC record embedding, which is a retrieval quality
metric that measures how well the KB matched the issue; and \texttt{gt\_keyword\_hit\_rate}: fraction of ground truth keywords appearing in the generated response, a model-free proxy for answer correctness, not subject to judge bias.

\begin{figure*}[t]
\centering
\includegraphics[width=0.9\linewidth]{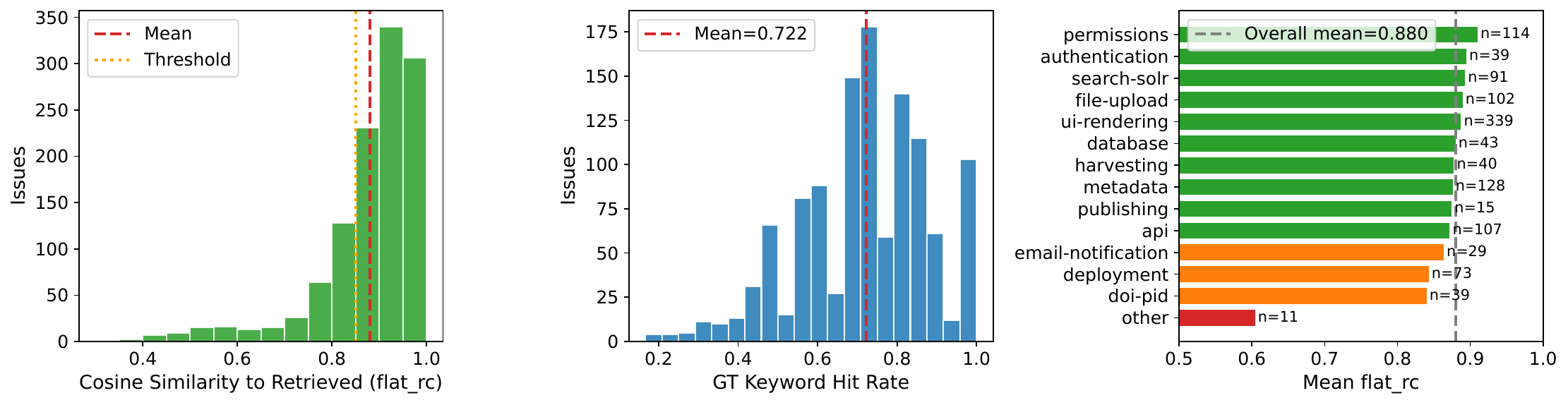}
\caption{Full benchmark retrieval quality ($n{=}1{,}172$ issues).
Left: \texttt{flat\_rc\_score} distribution (cosine similarity to retrieved RC triple);
55\% of issues score ${\geq}0.90$.
Center: GT keyword hit rate distribution (mean $= 0.722$).
Right: mean \texttt{flat\_rc\_score} by domain; 13 primary categories shown plus \textit{other} ($n{=}11$).}
\label{fig:full-benchmark}
\end{figure*}

\subsection{Knowledge Ablation}
We evaluate all four configurations on the full 1,172-issue benchmark
using the decomposed 4-dimension rubric scored by \texttt{claude-sonnet-4-6}.
All four configurations have complete agent responses for every issue;
judge scoring covers all 1,172 issues ($4{,}688$ total judge calls).
Table~\ref{tab:main-results} reports decomposed 4-dimension scores for all four
configurations.
95\% bootstrap confidence intervals (10{,}000 resamples) are included; no interval pair overlaps, confirming that all pairwise differences are statistically significant ($p{<}0.001$).

\omrag dominates on all diagnosis-relevant dimensions.
Diagnosis Accuracy of 0.931 is $+186\%$ over \cchunk (0.325) and
$+291\%$ over \czero (0.238).
Fix Correctness shows even larger gains ($+332\%$ over \cchunk).
The ordering \czero $<$ \cchunk $<$ \cdockg $<$ \omrag holds consistently
across all four dimensions and all 13 domain categories.
Without any retrieval, \czero achieves substantially lower diagnosis accuracy (0.238),
confirming that the retrieval contribution is essential, not incidental.
\czero's non-trivial Specificity (0.453) and low Groundedness (0.429) reflect
parametric Dataverse knowledge from LLM pre-training: the model names plausible
artifacts but cannot identify the actual fault pattern without retrieved precedents.

The two retrieval baselines each demonstrate one variant of the
retrieved but wrong failure mode identified in Section~\ref{sec:intro}.
\cchunk exhibits the \emph{structure mismatch} variant: it retrieves from
the correct source (operational issues) but unstructured chunking fragments
the symptom--root cause mapping, yielding the highest Groundedness (0.865)
yet low Diagnosis Accuracy (0.325)---the model faithfully uses whatever chunks
it receives but cannot reconstruct the complete causal chain.
\cdockg exhibits the \emph{source mismatch} variant: documentation-anchored
retrieval routes queries through a concept layer derived from system
specifications, achieving moderate DA (0.526) but lower Groundedness (0.684)
than \cchunk, indicating that the documentation-anchored path introduces
conceptual detours that dilute operational specificity.
Only \omrag, which combines operational source with structured triples,
addresses both failure modes simultaneously.

\subsection{Judge Reliability}
\label{subsec:judge}

Three studies validate that the \omrag advantage is measurement-stable rather than
an artifact of judge bias.

\noindent\textbf{Positional bias.}
On a 23-issue subset, we presented each response pair in normal and reversed order.
Winner agreement across orderings: $87\%$ ($20/23$ issues).

\noindent\textbf{Cross-judge validation.}
The same 23 issues were scored independently by \texttt{claude-sonnet-4-6} and
\texttt{gpt-5.2}. Spearman rank correlation on per-config total scores:
$r{=}0.943$ ($p{<}0.001$), indicating high cross-judge consistency.
\texttt{gpt-5.2} confirmed \omrag as winner or co-winner in all cases.

\noindent\textbf{Verbosity control.}
The decomposed 4-dimension rubric mitigates verbosity bias by scoring discrete
semantic properties (DA, FC, G, S) independently rather than overall response quality.

\noindent\textbf{Effect-size robustness.}
The three bias mitigation studies above were conducted on a 23-issue subset.
The observed DA gap between \omrag and \cchunk (${\approx}0.60$) substantially exceeds
the judge reliability variation documented by prior work~\cite{wang2023notfairevaluators,wu2024judgingjudges},
meaning that no plausible judge reliability error could reverse the ranking conclusion.

\noindent\textbf{Human verification.}
A 123-issue verification subset was manually reviewed, checking each judge
decision against the ground-truth root cause.
All decisions were confirmed as reasonable; the 8 non-\omrag outcomes (6.5\%) each
had identifiable explanations: 4 cases where \cdockg's concept layer matched a
specific technical detail more precisely, 2 trivial tasks (version bumps) where
zero-shot sufficed, and 2 cases where no configuration identified the correct root cause.

\subsection{Retrieval Quality at Scale}
\label{subsec:scale}

Figure~\ref{fig:full-benchmark} shows the model-free retrieval quality metrics
across 1,172 issues, providing independent corroboration without any LLM judge involvement.

\noindent\textbf{Retrieval quality is consistently high.}
Mean \texttt{flat\_rc\_score} = $0.880$ (stdev = $0.106$) across all 1,172 issues,
with 55\% of issues scoring ${\geq}0.90$ and 86\% scoring ${\geq}0.80$.
The distribution is left-skewed, indicating that the operational memory KB provides
high-quality symptom matches for the large majority of queries.

\noindent\textbf{No catastrophic domain gaps.}
GT keyword hit rate ranges from $0.682$ (file-upload) to $0.744$ (metadata)
across the 13 named domain categories, a variance of $0.062$, small relative to the gap
between \omrag and \cchunk on Diagnosis Accuracy ($+0.500$).
Even the weakest domain (\texttt{file-upload}) substantially outperforms
chunk-based retrieval on the evaluation metrics.

\subsection{Cost Analysis}
\label{subsec:cost}

The full 1{,}172-issue \omrag benchmark run cost \$65.22 in API calls
(mean \$0.056/issue), dominated by gpt-5.2 inference and claude-sonnet-4-6
judging. KB construction (embedding 1{,}457 records) cost approximately \$0.80.
The operational memory approach is thus economically viable for production
software systems with comparable issue volumes.

\subsection{Limitations}
\label{subsec:limitations}

Three factors bound the scope of these findings. First, the benchmark and KB
are derived from a single production system (Dataverse); while its multi-component
architecture and decade-long issue history are representative of many research-software
deployments, the magnitude of the source and structure effects may vary in systems
with different issue-tracking practices or shorter operational histories. Second,
although subset-based positional, cross-judge, and manual checks corroborate the ranking,
primary scoring relies on an LLM
judge rather than blind domain-expert annotation at full benchmark scale. Third, the
13 domain categories used for the breakdown in Section~\ref{subsec:scale} are assigned
by the same LLM extractor used for triple extraction; these labels affect only the
post-hoc analysis, not KB construction or retrieval, but an independently sourced
categorization could further strengthen the domain-level comparison. Extending the
benchmark to additional open-source systems is the most direct path toward establishing
the generality of operational memory for diagnostic RAG.

% ─────────────────────────────────────────────────────────────────────────────
\section{Related Work}
\label{sec:related}

\textbf{Graph RAG \& AIOps Retrieval.}
Existing Graph RAG methods improve retrieval through graph or hierarchical structures, including GraphRAG~\cite{edge2024graphrag}, HippoRAG~\cite{gutierrez2024hipporag}, LightRAG~\cite{guo2024lightrag}, and RAPTOR~\cite{sarthi2024raptor}. However, GraphRAG-Bench~\cite{xiang2025graphragbench} shows that graph structures provide limited benefits when answers reside in single records rather than requiring multi-hop reasoning, which is often the case for bug diagnosis. Recent AIOps systems retrieve historical incidents for diagnosis~\cite{goel2025earco,zhang2024icl,chen2024rcacopilot,wang2025arca}. While effective, they rely on proprietary datasets and unstructured retrieval, limiting reproducibility and cross-system comparison. In contrast, \omrag focuses on a different retrieval target: structured operational cases extracted from public GitHub issues, together with an open benchmark.

\textbf{Software Agents \& RAG Evaluation.}
Software engineering benchmarks such as SWE-bench~\cite{jimenez2024swebench} primarily evaluate patch generation rather than diagnosis. Prometheus~\cite{ma2025prometheus} augments code reasoning with graph structures but does not model operational failure history. We instead focus on diagnostic reasoning grounded in operational memory. Our evaluation follows prior work on RAG assessment~\cite{es2023ragas} and LLM-as-Judge~\cite{zheng2023mtbench}, while verifying correctness against ground-truth diagnoses.

\textbf{GenAI for Self-Adaptive Systems.}
Recent work identifies dynamic operational knowledge as a major challenge for GenAI-enabled self-adaptive systems~\cite{li2025genaisas}. MAPER~\cite{maper2026} extends MAPE-K with LLM-based reasoning. \omrag addresses the complementary problem: providing structured operational memory that accumulates resolved failures and grounds diagnosis in past experience.

\section{Conclusion}
\label{sec:conclusion}
We presented Operational Memory RAG (\omrag), which retrieves structured operational cases for software bug diagnosis. Across 1,172 real-world issues, operational history outperforms documentation and structured cases outperform unstructured chunks. \omrag operationalizes the Knowledge ($K$) component of MAPE-K as an episodic memory of past failures and repairs. Its deployment in \textbf{DIVA} (\url{https://dataverse.fiu.edu/ai/}), which has administered the FIU Dataverse production instance for over six months, demonstrates its practical utility. These results suggest that diagnostic RAG depends on retrieving the right operational experience.

\balance
\bibliographystyle{IEEEtran}
\bibliography{references}

\end{document}